\documentclass
[preprint,prb,superbib,superscriptaddress,byrevtex,showpacs,twocolumn,tightenlines]{revtex4}%
\usepackage{amsfonts}
\usepackage{amsmath}
\usepackage{amssymb}
\usepackage{graphicx}%
\begin{document}
\preprint{ }
\title{Comparison of $S=0$ and $S=1/2$ Impurities in Haldane Chain Compound, Y$_{2}%
$BaNiO$_{5}$}
\author{J. Das}
\affiliation{Department of Physics, Indian Institute of Technology, Mumbai 400076, India}
\author{A. V. Mahajan}
\affiliation{Department of Physics, Indian Institute of Technology, Mumbai 400076, India}
\author{J. Bobroff}
\affiliation{Laboratoire de Physique des Solides, UMR 8502, \ Universit\'{e} Paris-Sud,
91405 Orsay, France}
\author{H. Alloul}
\affiliation{Laboratoire de Physique des Solides, UMR 8502, \ Universit\'{e} Paris-Sud,
91405 Orsay, France}
\author{F. Alet}
\affiliation{Computational Laboratory and Theoretiche Physik, ETH Z\"{u}rich, CH-8092
Z\"{u}rich, Switzerland}
\author{E. S. S\o rensen}
\affiliation{Department of Physics and Astronomy, McMaster University, Hamilton, ON, L8S
4M1 Canada}

\begin{abstract}
We present the effect of Zn ($S=0$) and Cu ($S=1/2$)\ substitution at the Ni
site of $S=1$ Haldane chain compound Y$_{2}$BaNiO$_{5}$. $^{89}$Y NMR allows
us to measure the local magnetic susceptibility at different distances from
the defects.\ The $^{89}$Y NMR spectrum consists of one central peak and
several less intense satellite peaks. The central peak represents the chain
sites far from the defect.\ Its shift measures the uniform susceptibility,
which displays\ a Haldane gap $\Delta\ \approx100$ K and it corresponds to an
AF\ coupling $J\approx260$ K\ between the near-neighbor Ni spins.\ Zn or Cu
substitution does not affect the Haldane gap. The satellites, which are evenly
distributed on the two sides of the central peak, probe the antiferromagnetic
staggered magnetization near the substituted site. The spatial variation of
the induced magnetization is found to decay exponentially from the impurity
for both Zn and Cu for $T>50$ K.\ Its extension is found identical for both
impurities and corresponds accurately to the correlation length $\xi
(T)$\ determined by Monte Carlo (QMC) simulations for the pure compound. In
the case of non-magnetic Zn, the temperature dependence of the induced
magnetization is consistent with a Curie law with an \textquotedblleft
effective\textquotedblright\ spin $S=0.4$ on each side of Zn. This staggered
effect is quantitatively well accounted for in all the explored range by
Quantum Monte Carlo computations of the spinless-defect-induced magnetism. In
the case of magnetic Cu, the similarity of the induced magnetism to the Zn
case implies a weak coupling of the Cu spin to the nearest- neighbor Ni
spins.\ The slight reduction of about $20-30$\% of the induced polarization
with respect to Zn is reproduced by QMC computations by considering an
antiferromagnetic coupling of strength $J^{\prime}=0.1-0.2$ $J$\ between the
$S=1/2$ Cu-spin and nearest-neighbor Ni-spin. Macroscopic susceptibility
measurements confirm these results as they display a clear Curie contribution
due to the impurities nearly proportional to their concentration.\ This
contribution is indeed close to that of two spin-half for Zn
substitution.\ The Curie contribution is smaller in the Cu case, which
confirms that the coupling between Cu and near-neighbor Ni is antiferromagnetic.

\end{abstract}
\eid{identifier}
\date{26 October, 2003}
\startpage{1}
\endpage{ }
\maketitle

\section{\textbf{INTRODUCTION}}

One and two-dimensional Heisenberg antiferromagnetic ($1D$ and $2D$ HAF) spin
systems are presently attracting strong interest due to the variety of ground
states that these systems exhibit : spin-gap, spin-Peierls, RVB,
superconductivity, etc. Effects of substitution in the HAF
half-integer\cite{a} and integer-spin chains,\cite{13} ladders,\cite{c} and
planes\cite{d} have been studied both experimentally and theoretically to
understand such ground states and their elementary excitations.

Among these systems, one-dimensional $S=1$ HAF chains have attracted immense
interest after Haldane\cite{1} conjectured that in case of integer-spin
chains, there is a finite energy gap between the ground state and first
excited state. The ground state of this system can be well described by
Valence Bond Solid (VBS) model.\cite{2} This model suggests that each $S=1$
spin can be considered to be a symmetric combination of two $S=1/2$ spins.
With periodic boundary conditions, each $S=1/2$ forms a singlet with the
nearest-neighbor (n.n.) belonging to the adjacent site. Introducing a
non-magnetic impurity, i.e. open boundary conditions, is equivalent to
removing one $S=1$. This is expected to give rise to two $S=1/2$ at each end
of the chain.\cite{3}

Y$_{2}$BaNiO$_{5}$ (YBNO), an $S=1$ one-dimensional HAF is a well established
Haldane gap compound. YBNO can be described by the space group $I_{mmm}%
$.\cite{4} All the Ni$^{2+}$ ions are located at the center of highly
compressed corner shared NiO$_{6}$ octahedra and form Ni - O - Ni chains along
the $a$- axis (see Fig. 1 (a)). These octahedra are flattened along the chain
axis and the Ni - O distances can be divided in two groups (Ni -- O$1$
$\approx$\ $1.885$ \AA \ and Ni -- O$2$ $\approx$\ $2.186$ \AA ). Intra-chain
Ni -- Ni distance is equal to $3.77$\AA \ (Ni$^{2+}$ ions interact via the
oxygen ion), while the smallest inter-chain Ni - Ni distance is close to $6.6$
\AA . Also, since these chains are separated by Ba and Y cations, the physical
properties are expected to be predominantly one-dimensional ($1D$). The
superexchange coupling $J$ estimated from magnetic susceptibility is $J$
$\approx$\ $285$ K.\cite{5} The magnitude of the Haldane gap $\Delta$
$\approx$\ $100$ K has been obtained from magnetic susceptibility
measurements.\cite{5,b,7} Inelastic neutron scattering experiments (INS) also
reveal the existence of a gap $\Delta$ $\approx$\ $100 $ K,\cite{5,8} between
the ground state and first excited state and $|J_{\perp}/J|<5\times10^{-4}$,
where $J_{\perp}$ and $J$ are the inter-chain and intra-chain coupling
constants, respectively.\cite{9} This proves the strong $1D$ character of the
Ni chains. $^{89}$Y NMR experiments on polycrystalline YBNO,\cite{10} also
show the presence of a gap\textbf{\ }$\Delta$ $=$\ $80\pm5$ K\textbf{\ }and an
anisotropic Knight shift which is responsible for the asymmetry in the NMR spectra.

In-chain Zn$^{2+}$, Mg$^{2+}$ ($S=0$) substitutions at the Ni$^{2+}$ ($S=1$)
site of YBNO\cite{13,20,22,24} have been attempted. Bulk susceptibility
measurements confirm that non-magnetic substitutions break the chains and give
rise to chain-end free spins.\cite{22} Low-temperature specific heat
measurements\cite{20} on different doped samples have been quantitatively
explained by an effective model,\cite{23} which describes the low energy
spectrum of non-interacting open $S=1$ chain. Tedoldi \textit{et al.}\cite{24}
carried out $^{89}$Y NMR experiments in $5$\% and $10$\% Mg doped YBNO for
temperatures $T>$ $\Delta$. They observed multiple-peak $^{89}$Y NMR spectra
at different $T$ and determined the $T$-dependence of the staggered
magnetization near the impurity from the shift of these peaks at different
$T$. From the site-dependence of the staggered magnetization, the
characteristic length scale of its decay was found\ to be similar to the
correlation length $\xi(T)$ of the pure $S=1$ HAF chain. At lower
temperatures, the NMR spectra were found to broaden significantly and hence
various features could not be resolved. These results prove that the
non-magnetic impurity-induced effects reveal the intrinsic characteristics of
the correlation functions of these systems, both their exponential shape and
their extension. Such results were quantitatively explained by Monte Carlo
\cite{25,26,29} and Density Matrix Renormalization Group simulations.
\cite{29,27,28}

In this paper, we propose to extend such local studies to other impurities,
both non-magnetic and magnetic and to greatly improve the quantitative aspects
of these results. We, therefore, present a study of the susceptibility and
$^{89}$Y NMR spectra on Y$_{2}$BaNi$_{1-x}$Zn$_{x}$O$_{5}$ (YBNO:Zn) and
Y$_{2}$BaNi$_{1-x}$Cu$_{x}$O$_{5}$ (YBNO:Cu) ($0$ $\leq$ $x\leq$ $0.02$).
Dilute concentration of substitutions were preferred in an attempt to observe
well resolved impurity-induced effects on the $^{89}$Y NMR spectra far from
the impurities and in a temperature regime $T<$ $\Delta$. We detail the
experimental conditions in Sec. II and NMR and bulk susceptibility
measurements on YBNO:Zn and YBNO:Cu will be presented in Sec. III. Zn and Cu
substitutions are found to induce low-$T$ Curie-terms in the bulk
susceptibility, while $^{89}$Y NMR spectra show a multi-peak structure. The
multiple peaks are associated to an impurity-induced short-range staggered
magnetization which appears very similar for both Zn and Cu. As discussed in
Sec. IV, our results on YBNO:Zn prove the universality of the effects of
spinless (Mg and Zn) impurities in the $S=1$ chain. These results are
quantitatively reproduced by our QMC\ simulations. Effects of an $S=1/2$
impurity (Cu) are contrasted with those of Zn substitution at various
temperatures.\ Here again, QMC\ simulations fit the data well and enable us to
deduce the magnitude of the exchange coupling between Cu and n.n. Ni, which is
found weak and antiferromagnetic. Finally, the contribution of impurity
induced magnetization to the total susceptibility is extracted from NMR for
the first time and has been compared to the macroscopic (SQUID) measurements
taken on the same set of samples. All results can be well explained in a
common VBS-like framework both for Zn and Cu.

\section{\textbf{EXPERIMENTAL DETAILS}}

Polycrystalline samples of Y$_{2}$BaNi$_{1-x}$Zn$_{x}$O$_{5}$ and Y$_{2}%
$BaNi$_{1-x}$Cu$_{x}$O$_{5}$\ ($0$ $\leq$ $x\leq$ $0.02$) were prepared by
standard solid state reaction techniques. Stoichiometric mixtures of pre-dried
Y$_{2}$O$_{3}$, BaCO$_{3}$, NiO, ZnO, and CuO ($>99.9$\% pure) were fired in
air at $1200%
{{}^\circ}%
$C. Single phase samples were obtained after several rounds of grinding,
pelletization, and firing. In an attempt to reduce the parasitic Curie term in
the susceptibility, as discussed in Sec. IIIA, we fired the samples in vacuum
at $400%
{{}^\circ}%
$C, in order to reduce the excess oxygen present in the sample. X-ray powder
diffraction was carried out using a PHILIPS PW $1729$ powder diffractometer,
using Si as\ an internal standard. A Cu target was used in the diffractometer
with $\lambda_{av}$ = $1.54182$ \AA . Lattice parameters were determined from
the diffraction pattern using a least square fit method. The lattice
parameters for the\textquotedblright pure\textquotedblright\ YBNO sample were
found to be $3.763(1)$ \AA , $5.762(1)$ \AA ,\ and $11.334(3)$ \AA \ along
$a$-, $b$-, and $c$- axes respectively, which are in agreement with the
earlier results.\cite{4} The lattice parameters remain almost unchanged after
Zn and Cu substitution. The nearly identical effective ionic radii\cite{30} of
Ni$^{2+}$, Zn$^{2+}$ and Cu$^{2+}$, together with the dilute level of doping
may be the cause of the unchanged lattice parameters.

Magnetization $M$ measurements were performed as a function of temperature $T
$ ($1.8$ K $\leq T$ $\leq$ $300$ K) and an applied field $H$ ($0$ $\leq$ $H $
$\leq$ $50$ kGauss) using a Superconducting Quantum Interference Device
(SQUID) magnetometer.

NMR measurements were performed by standard pulsed NMR techniques in an
applied field $H_{0}$= $70$ kGauss at different temperatures ($50$ K $\leq$
$T$ $\leq$ $325$ K). The typical width of the applied $\pi/2$ pulse was
$t_{W}=6$ $\mu s$. The line shape was obtained by Fourier Transform (FT) of
the NMR echo signals. The $^{89}$Y NMR Knight shift $\ K(T)=$ $\left[
\nu(T)-\nu_{ref}\right]  /\nu_{ref}$ was measured with respect to a standard
YCl$_{3}$ solution ($\nu(T)$ is the resonance frequency of Y$_{2}$BaNiO$_{5} $
at different temperatures and $\nu_{ref}=14594.15$ kHz is the resonance
frequency of YCl$_{3}$ in our experiment). At higher temperatures ($120$ K
$\leq T\leq300$ K), the spectral width of the $\pi/2$ pulse was sufficient to
reliably obtain the full spectrum. However, at lower temperatures ($T<120$ K),
where the multiple resonance lines are spaced far apart in frequency, we
varied the irradiation frequency to obtain the full spectrum as the spectral
width of the single $\pi/2$ pulse was insufficient.

The Quantum Monte Carlo simulations were performed using the Stochastic Series
Expansion method,\cite{alet1,alet2} with generalized directed loop
techniques.\cite{alet3, alet4, alet5} Calculations of magnetization profiles
were done for long runs of about $2.10^{8}$ Monte Carlo steps for an isotropic
$S=1$ Heisenberg model, with an imposed magnetic field of about $70 $ kGauss,
corresponding to the field used in the NMR experiments.

\section{\textbf{RESULTS}}

\subsection{Bulk Susceptibility}

Susceptibility\ $\chi(T)$ of Y$_{2}$BaNi$_{1-x}$Zn$_{x}$O$_{5}$ and Y$_{2}%
$BaNi$_{1-x}$Cu$_{x}$O$_{5}$ ($0$ $\leq$ $x\leq$ $0.02$) was measured as a
function of $T$ in an applied field of $H=5$ kGauss\ (Fig. 2).\ We checked for
the possible presence of spurious ferromagnetic impurities by measuring $M$ vs
$H$ isotherms at different temperatures in order to separate the linear
(paramagnetic) and non-linear (ferromagnetic) contributions. The ferromagnetic
contribution corresponds to about $2$ ppm of ferromagnetic Fe impurity which
amounts to at most $0.2$\% of the susceptibility. It is, hence, neglected in
our analysis. The paramagnetic susceptibility for the \textquotedblleft
pure\textquotedblright\ sample is in rough agreement with the results
published earlier.\cite{10,12} The susceptibility decreases\textbf{\ }%
as\textbf{\ }$e^{-\frac{\Delta}{T}}/\sqrt{T}$ with decreasing temperature, as
a consequence of the Haldane gap.\cite{AffleckA} At lower temperatures ($T<50$
K), there is a Curie-Weiss like$\ C/(T+\theta$) upturn in the susceptibility.
This parasitic Curie term in the \textquotedblleft pure\textquotedblright%
\ sample might presumably result from various sources such as natural chain
breaks, presence of Ni$^{3+}$ (as a result of small oxygen non-stoichiometry),
and extrinsic paramagnetic phases present in the sample. A substantial
reduction in this low-$T$ contribution to the susceptibility (i.e. reduction
in the Curie term $C$) was achieved by vacuum annealing the powders at $400%
{{}^\circ}%
$C, as evident from the inset of Fig. 2 (a).\ A similar reduction in the
parasitic Curie-term in a $S=1/2$ chain compound, Sr$_{2}$CuO$_{3},$ was
observed earlier by Ami \textit{et al.}\cite{31} after annealing in inert
atmosphere. Reduction in the Curie term in our samples might be due to removal
of excess oxygen leading to a reduction in the amount of Ni$^{3+}$ due to the
above process. A discussion about intrinsic defects, effect of vacuum
annealing and the implications on SQUID\ and NMR measurements is carried out
in Sec. IVD. All measurements with the substituted samples reported hereafter
are on the vacuum annealed samples. In case of the substituted samples, the
susceptibility at lower temperatures increases with increasing dopant content
(shown in Fig. 2 (a,b), Inset), which indicates an increase in the Curie-term
for the substituted samples.

\subsection{$^{89}$Y NMR}

In YBNO, each $^{89}$Y has two n.n. Ni$^{2+}$ from one chain and two
next-nearest-neighbor (n.n.n.) Ni$^{2+}$ from two different chains (Fig. 1 (a,
b)). However, $^{89}$Y is actually coupled to the Ni$^{2+}$ spins belonging to
n.n.n. chains, not to the n.n. one.\ Indeed, there is no appreciable exchange
between Y and Ni $3d$ or oxygen $p$ orbitals belonging to n.n. chain. In
contrast, there is an appreciable overlap between the n.n.n.-oxygen $p_{\pi}$
orbital with the $5s$- orbital of Y$^{3+}$ and hence with the n.n.n. Ni - O -
Ni chains.

We performed $^{89}$Y NMR on the ``pure''\ and substituted samples at
different temperatures. In case of the annealed and non-annealed
''pure''\ samples, we observed one asymmetric central peak (inset: Fig. 3)
shifted from $\nu_{ref}$ in the Fourier transformed spectrum, together with
less intense peaks (satellites) on both sides of the central peak (see Fig.
3). The shift in the $^{89}$Y NMR peak of the ''pure'' YBNO arises from the
coupling of Y nuclei to Ni$^{2+}$ spins of two different n.n.n. chains. Since
our measurements are on randomly aligned polycrystalline samples, asymmetric
lineshape of the central peak (mainline) in the annealed sample corresponds to
a powder pattern due to an asymmetric hyperfine coupling tensor $A_{hf}%
$\cite{10} and anisotropic susceptibility.\cite{7} The vacuum annealing
procedure leads to a large narrowing of the mainline and a decrease of the
satellites intensities.\ The amount of defects corresponding to these
satellites can be determined by measuring their area. Using an asymmetric
Lorenzian fit, the fractional area of one of these satellites decreases\ from
$3.2$\% to $0.8$\% after annealing. These satellites in ''pure''\ YBNO
presumably originate from the in-chain defects, which will be discussed in
details in Sec. IVD.

In substituted compounds, the shift of the mainline for the substituted
samples, i.e. Knight shift $K(T)$, remains unchanged from the ''pure''\ sample
at different temperatures (Fig. 4) and compares well with the results
published earlier.\cite{10} This implies that the Haldane gap remains the same
after impurity substitution. In contrast, the introduction of impurities
(Zn/Cu) at the Ni-site of the chain, results in a change in the local
environment of the impurity. This change is evident by the appearance of
satellites on both sides of the mainline. For example, in Fig. 5, we compare
the NMR spectra of ``pure'', $0.5$\%, and $2$\% Zn substituted samples at
$120$ K (Fig. 5 (a)) and $0.5$\% and $1$\% Cu substituted samples at $300$ K
(Fig. 5 (b)). The intensity of the impurity-induced satellites increases with
increasing impurity concentration keeping their shift from the mainline
constant.\ This establishes that the satellite peaks are a direct consequence
of substitutions and their positions are independent of their concentration ,
i.e. of the length of the chains. Furthermore, fractional intensities of these
Cu and Zn-induced satellites are significantly larger than the weak satellites
observed in the annealed ``pure''\ sample. Hence, they can safely be
attributed to Zn or Cu neighbors. Each of these satellites represents $^{89}$Y
nuclei at a given distance from the impurity.\ Since one Y nucleus is coupled
to two different n.n.n. chains (see Fig. 1 (b)), for a concentration $x$ of
the impurity, there should be $4x$ Y-nuclei contributing to each satellite.
Using an asymmetric Lorenzian fit (Fig. 6), we extracted the fractional areas
of one satellite peak to the whole spectrum (outermost satellite on the right
side of the mainline, marked by $1$ in Fig. 6).\ For a nominal concentration
$x=2$\% of Zn or Cu, our fit leads to an effective concentration of
$(1.9\pm0.1)$\%\ for Zn and $(1.8\pm0.2)$\% for Cu. This good agreement
between the nominal stoichiometry and the actual in-chain number of defects
shows that the substitution amount is well controlled here.

Comparison of Cu and Zn $^{89}$Y NMR spectra done in Fig. 7 shows that Cu
induces a staggered magnetization like Zn.\ But the difference in the shifts
of the NMR satellites shows that the amplitude of the induced magnetization
near Cu is smaller than that for Zn. Representative $^{89}$Y spectra are shown
at different temperatures in Figs. 8 (a) and 8 (b) for $2$\% Zn and $0.5$\% Cu
substituted samples respectively. The Ni$^{2+}$ ions closest to the impurity
are the most affected by the substituent and give rise to the most shifted
satellite indexed as the $1^{st}$ neighbor ($1$ in Figs. 8 (a) and 8 (b)).
Further neighbors are numbered in accordance with the decreasing shifts of
satellites with respect to mainline. Satellites with decreasing shifts
alternate on each side of the mainline. This shows that the corresponding Ni
sites have a magnetization, which alternates in sign in addition to the
non-perturbed magnetization represented here by the mainline shift.\ The
appearance of such induced staggered magnetization near the defects is indeed
expected by theoretical computations,\cite{29,33} as will be quantitatively
detailed hereafter.

In the case of Zn, the nearest Y nuclei to the impurity (marked by $0$ in Fig.
1\ (b))\ couple to both Zn itself and one undisturbed chain. Therefore, its
corresponding NMR line should be shifted by half of that of the mainline with
respect to $\nu_{ref}$,\ as Zn is non-magnetic and hence does not contribute
to $K$. Furthermore, this satellite should be half the intensity of the other
satellites, from its specific position. In fact, this satellite has been
observed clearly in Fig. 8 (a) where it is labelled as ``$0$'': it is half as
shifted as the mainline and half as intense as the other satellites. In the
case of YBNO:Cu, such a peak is not observed as seen in Fig. 8 (b).\ This is
not surprising as the $S=1/2$ Cu-spin will now contribute to its shift, which
should now become large, and would be severely affected by a large
spin-lattice relaxation rate. Therefore, the absence of such a peak is a clear
local evidence that Cu has indeed successfully substituted at a Ni site.

The shifts of the observed satellites were measured for different temperatures
and concentrations. We note the shifts of the $i^{th}$ satellite as $\delta
\nu_{i}=\nu_{i}-\nu_{main}$, where $\nu_{i}$ and $\nu_{main}$ are frequencies
corresponding to its center of gravity and that of the mainline respectively.
Due to the asymmetric powder pattern shapes, consideration of the centre of
gravity for each peak yields a better description of the impurity-induced
effects in the chain. In Fig. 9 (a), we present these shifts of the satellites
($\delta\nu_{i}$) at various temperatures for the spinless Zn impurity. On the
same figure, we plot as well the data published earlier for the Mg impurity,
which is found identical to that of Zn.\cite{24}\ Therefore, the induced
magnetization by both Zn and Mg is quantitatively the same. In the present
study, the use of samples with smaller impurity content and less natural
defects allowed us to largely improve the resolution and to detect more
satellites in a wider temperature range. A similar plot is shown in Fig. 9 (b)
for YBNO:Cu. These measurements allow us to get a direct measure of the
amplitude and spatial shape of the magnetization induced by both non-magnetic
and magnetic impurities, as will be discussed later.

\section{DISCUSSION}

\subsection{Determination of the parameters of the Haldane Hamiltonian of the
pure system}

The bulk susceptibility $\chi(T)$ in the \textquotedblleft
pure\textquotedblright\ YBNO contains the contribution from some unknown
defects and extrinsic paramagnetic phases in addition to the intrinsic gap
contribution. On the other hand, in $^{89}$Y NMR spectra, defects and
intrinsic behavior can be easily distinguished since defects broaden the
mainline or induce satellites, whereas the intrinsic behavior is measured by
the mainline shift. Hence, the $T$--dependence of the \textquotedblleft gap
term\textquotedblright\ is best reflected by the temperature dependence of the
NMR Knight shift $K(T)$.

Indeed, the Knight shift $^{89}K(T)$ of the central line is expected to vary
linearly with the uniform susceptibility $\chi_{u}(T)$ through
\begin{equation}
K(T)=\frac{2A_{hf}}{\mu_{B}N_{A}}\chi_{u}(T)+K(0)
\end{equation}

Here, $A_{hf}$\ is the $^{89}$Y - Ni Hyperfine coupling constant and is
independent of temperature, $N_{A}$\ is the Avogadro number, and $K(0)$ is the
chemical shift independent of temperature. The NMR shift $K(T)$ results from
two n.n.n. chains, while $\chi_{u}(T)$ represents the intrinsic susceptibility
of one $S=1$ chain, which explains the factor $2$. As reported in Fig. 4, the
Knight shift $K(T)$ of the central peak decreases with decreasing $T$ and
remains unchanged for all the samples at the corresponding$\;T$. The observed
exponential decrease in $K(T)$ at low-$T$ is direct evidence of the presence
of the Haldane gap. Our measurements on different samples at different
temperatures match well with earlier results of Shimizu \textit{et al.} for
\textquotedblright pure\textquotedblright\ YBNO\cite{10} (as shown in Fig. 4).
This indicates that the gap remains unchanged for impurity substituted
samples. I.\ Affleck showed that the low temperature part of the
susceptibility should follow $\sqrt{\Delta/T}e^{-\frac{\Delta}{T}}%
$.\cite{AffleckA} Previous studies used this model to fit the low-$T$ part of
both NMR and macroscopic susceptibility measurements in order to extract the
Haldane gap and the superexchange coupling $J$. This model also fits our data,
as shown by a solid line in Fig. 4 for the Cu$0.5$\% compound, leading to
$K(0)=340$ ppm and $\Delta=(88\pm7)$ K, which is in good agreement with the
results published earlier.\cite{9,10}

However, we propose here a more rigorous way to determine $J$. We used QMC
simulations to directly fit our data in the full temperature range. We have
calculated theoretically, by QMC simulations, the uniform susceptibility
$\chi_{u}$\ versus $T/J$ i.e. the susceptibility of the infinite $S=1$\ chain
theoretically, by QMC simulations. The Knight shift $K$ obtained from our NMR
experiments has then been plotted versus this $\chi_{u}$ for different $J $
values in Fig. 10 with $T$ as an implicit parameter. The $T\rightarrow0$ value
of $K$ is constrained by Shimizu's NMR measurements,\cite{10} which allowed to
accurately determine $^{89}K(T\rightarrow0)=370\pm10$ ppm, a value compatible
with our measurements. Using this constraint, a good linear fit is obtained
for $J=250\pm10$ K (shown by the solid line). This proves that the uniform
susceptibility measured by NMR is well accounted by QMC\ simulations up to
$T=350$ K. This procedure of determining $J$ allows for a much sharper
constraint on the determination of $J$ than the previous determinations. We
believe that this procedure enables the determination of $J$ more reliable
than before. The corresponding theoretical Haldane gap\cite{33} would be
$\Delta=0.41J=102\pm4$ K, a value compatible with our first rough estimate and
with the previous experiments.\cite{7} This fitting method also allows a
determination of the hyperfine coupling $A_{hf}$ independent of any
macroscopic susceptibility measurement. We performed similar fitting for all
substituted and pure YBNO shifts plotted in Fig. 10.\ This leads to
$A_{hf}=(4.7\pm0.2)$ kGauss/$\mu_{B}$.\ This value should be more reliable
than earlier determinations done by Shimizu \textit{et al.} \cite{10}
($A_{hf}=4.9$ kGauss/$\mu_{B}$) and by Tedoldi \textit{et al.}\cite{24}
($A_{hf}=6.5$ kGauss/$\mu_{B}$), which relied too much on the actual
measurements of the susceptibility of the pure compound, as this measurement
is always hampered by the large low-temperature Curie contributions.

\subsection{Impurity induced staggered magnetization}

\subsubsection{Spatial dependence of the induced staggered magnetization}

In Figs. 11 (a) and 11 (b), we depict the site-dependence of $|\delta\,\nu
_{i}(T)|$ at different temperatures for YBNO:Zn and YBNO:Cu, respectively, on
a logarithmic scale. At a given $T$, the observed linear behavior shows that
$|\delta\nu_{i}(T)|$\ varies exponentially as a function of distance from the
impurity.\ In fact, the shifts can be fitted at all temperatures for both Cu
and Zn by:
\begin{equation}
|\delta\nu_{i}\,(T)|=|\delta\,\nu_{1}(T)|e^{-\frac{(i-1)}{\xi_{imp}(T)}}%
\end{equation}
Here $\xi_{imp}(T)$ is the extension of the impurity-induced staggered
magnetization.\ The values of $\xi_{imp}(T)$ extracted from our measurements
are represented on Fig. 12 and compared with those obtained by Tedoldi
\textit{et al}. in the case of Mg substitution.\cite{24}\ It clearly appears
that Cu, Zn and Mg induce a staggered magnetization with the same spatial
extent.\ Furthermore, this spatial extent $\xi_{imp}(T)$ is found almost
identical to the correlation length computed for an infinite $S=1$ chain with
no defects.\cite{33} The experimental values of $\xi_{imp}(T)$\ are
systematically found about $5$\%\ higher than the values of $\xi(T)$%
\cite{33}$\;$but this discrepancy is still within experimental accuracy (e.g.
at $200$ K, $\xi_{imp}=2.5\pm0.2$, while $\xi=2.3$).\ The lowest temperature
data, which were only taken in the case of Cu impurities, depart a little bit
more from the value of $\xi(T)$. The good agreement found shows that for both
$S=0$ and\ $S=1/2$\ impurities, the characteristic length of the staggered
magnetization near defects reveals indeed the intrinsic $\xi(T)$ of the
infinite $S=1$ chain,\ in the explored temperature range.

\subsubsection{Comparison with QMC for Zn and Cu}

For both $S=0$ (Zn)\ and $S=1/2$ (Cu) impurity, we compared our experimental
results to those predicted by QMC calculations for the local susceptibility
around impurities. As discussed earlier, the total magnetization $M_{i}$\ at
the $i^{th}$ site in the chain is the sum of the uniform magnetization and the
impurity-induced magnetization i.e. $M_{i}=g\mu_{B}<S_{z}^{i}>=g\mu_{B}%
(<S_{z}>_{u}+\delta<S_{z}^{i}>)$. We can determine directly, from the
NMR\ data for $\delta\nu_{i}\;$and the mainline shift, the ratio of the
impurity-induced staggered magnetization and uniform magnetization
$\delta<S_{z}^{i}>/<S_{z}>_{u}$, using the value of $K(0)=340$\ ppm deduced
from Fig 4\ and 10. This experimental quantity could be directly compared with
that computed by QMC\ simulation for the spatial dependence of the spin
polarization at chain ends.

In case of the $S=0$ impurity, we performed such QMC calculations considering
$J=230$ K, $250$ K and $270$ K at $T=100$ K. The computed impurity-induced
staggered magnetization decays exponentially away from the impurity as in the
experimental results, following Eq. 2 (Fig. 13 (a)). The computed extension
$\xi_{imp}$ matches well with the previous results\cite{33} considering
$J=270$ K. Further, we found that the experimental ratio of the
impurity-induced staggered magnetization and uniform magnetization match well
with QMC for $J=270$ $\pm10$ K as shown in Fig. 13 (a). This value of $J$ is
consistent with the one independently determined from the uniform
susceptibility far from the defect $J=250\pm10$ K.\ This proves the coherence
of our analysis, as the same coupling $J$ can account both for the
susceptibility of the ``pure''\ compound, and for the shape, amplitude and
extension of the induced magnetization near spinless-defects. The
determination of $A_{hf}$ enables us to convert the NMR\ shifts into absolute
magnetization values. Here, we have calculated $<S_{z}^{i}>_{Zn}$, from our
experimental data using the relation $<S_{z}^{i}>=2\pi(2\nu_{i}-(\nu_{ref}%
+\nu_{main}))/2\gamma gA_{hf}$. Here, $\gamma/2\pi=2.0859$ MHz/Tesla is the
$^{89}$Y gyromagnetic ratio and $g$\ is the Land\'{e} $g$-factor. The
temperature dependence of $<S_{z}^{i}>$ is also well fitted by the
QMC\ simulations as exemplified in Fig.13 (b).

In case of the $S=1/2$\ substitution, theoretical calculations were done
earlier to predict the behavior of $S=1/2$\ impurity in $S=1$%
\ chain.\cite{27,37,38}\ The $S=1/2$ impurity is modelled by considering that
its coupling $J^{\prime}$ to its near-neighbor Ni can be different from $J$.
We have performed QMC calculations for various $J^{\prime}$ ranging from $-J$
to $+J$. The magnetic impurity is found to induce a staggered magnetization as
in the spinless case.\ Furthermore, its extension is found identical for
temperatures higher than $J^{\prime}$. These results are then in perfect
agreement with our data. For $T>J^{\prime}$, the only changes, which begin to
appear when comparing the $S=1/2\;$and $S=0$ induced effects in QMC results,
is the amplitude of the staggered magnetization. We found that for a
ferromagnetic coupling of the impurity to the n.n. Ni ($J^{\prime}<0$), the
staggered magnetization on the $1^{st}$, $2^{nd}$\ and further neighbors is
larger than that for a non-magnetic impurity.\ In contrast, an
antiferromagnetic coupling ($J^{\prime}>0$) leads to a smaller magnetization.
The experimentally observed $\delta\nu_{i}$\ for YBNO:Cu is less than that for
YBNO:Zn ($J^{\prime}=0$) at the corresponding temperature, as exemplified in
Fig. 7. This strongly suggests that the Cu - Ni coupling is then
antiferromagnetic. A quantitative comparison between QMC results and
experiments enables us to deduce $J^{\prime}$.\ In Fig.14, we compared the
experimental ratio $(\delta\nu_{i})_{Cu}/(\delta\nu_{i})_{Zn}$\ to that
theoretically obtained for $<\delta S_{z}^{i}>_{Cu}/<\delta S_{z}^{i}>_{Zn}$
for different values $J^{\prime}/J$.\ The experimental and theoretical results
match well for $0.1J\leq J^{\prime}\leq0.2J,\,$which\ allows to deduce an
estimate $J^{\prime}=(0.15\pm0.05)J$.

\subsubsection{Determination of the total induced susceptibility}

The magnetization induced by impurities can be written as $\delta M_{i}%
=g\mu_{B}\delta<S_{z}^{i}>$, where $\delta<S_{z}^{i}>=2\pi\delta\nu_{i}/\gamma
gA_{hf}$.\textbf{\ }This conversion enables us to deduce from NMR, the
contribution of substituted impurities $\chi_{imp}^{NMR}$ to the macroscopic
susceptibility. As the impurity induced magnetization decays exponentially as
a function of site on either side of the impurity in the $1D $ chain,
$\chi_{imp}^{NMR}$ can be expressed as:%

\begin{equation}
\chi_{1}(T)=\frac{\delta M_{1}(T)}{H_{0}}%
\end{equation}

\begin{equation}
\chi_{imp}^{NMR}(T)=\frac{2xN_{A}\chi_{1}(T)}{100}%
{\displaystyle\sum\limits_{i}}
(-1)^{i-1}e^{-\frac{i-1}{\xi_{imp}(T)}}%
\end{equation}

Here, $\chi_{_{1}}(T)$ is the impurity induced susceptibility on the $1^{st} $
neighbor to the impurity and $N_{A}$ is the Avogadro number, $x$ is the amount
of impurities per Ni in \%. The variation of $\chi_{imp}^{NMR}(T)$, shown in
Fig. 15, can be fitted to $\chi_{imp}^{NMR}(T)=C^{NMR}/T$. From the fit, we
get $C_{Zn}^{NMR}=(5.6\pm0.6)\times10^{-3}$ cm$^{3}$ K/mole/\% Zn and
$C_{Cu}^{NMR}=(4.7\pm0.5)\times10^{-3}$ cm$^{3}$ K/mole/\% Cu. These Curie
terms lead to the corresponding spin value $S$ on each side of the impurity
through $C=2N_{A}g^{2}S(S+1)\mu_{B}^{2}/3k_{B}$.\ For Zn, one finds an
impurity-induced ``effective spin''\textbf{\ }$S=0.4\pm0.06$ on either side of
the impurity in the chain, which is close to the VBS predicted $S=0.5$. In
case of YBNO:Cu, as $\chi_{Cu}^{NMR}$\ is also found to follow a Curie
behavior, the impurity-induced susceptibility on either side of the impurity
behaves almost as a free spin in the explored $T$-range. The lower value of
$C_{Cu}^{NMR}$\ than $C_{Zn}^{NMR}$\ is well accounted for, hereagain, by
QMC\ simulations, if the Cu-spin couples weakly and antiferromagnetically to
the nearest-neighbor Ni-spin.

\subsection{Temperature dependence of the macroscopic susceptibility}

As already stressed, the macroscopic susceptibility presented in Fig. 2
results from distinct contributions: the uniform contribution $\chi_{u}%
(T)\;$which exhibits the Haldane gap which\ decreases exponentially with
decreasing temperature; the intrinsic defects such as Zn or Cu, and the other
unknown defects and paramagnetic species, which are already present in the
pure compound. The latter two contributions mainly give rise to a Curie-like
magnetization in the low-temperature region. We plotted these different
contributions in the case of pure\ YBNO\ in Fig. 16. Note that the total
susceptibility not only consists of the Haldane term and a Curie term, but
also includes a residual non-Curie term, which could be due to some defects
that are not purely paramagnetic. This small residual non-Curie term exists in
all our samples, as evidenced in Fig. 17 and does not show any obvious
connection with the substitution. It, therefore, limits any quantitative
analysis of the substitution effect on macroscopic susceptibility.
Furthermore, the large contribution of the Curie term present at low-$T$ even
in the pure compound also severely limits any analysis, contrary to NMR
experiments for which the Zn and Cu effects could be isolated. However, at low
temperatures, the total susceptibility can be tentatively modelled by:%

\begin{equation}
\chi_{SQUID}(T)=\chi_{0}+\frac{C^{SQUID}}{T+\theta}+\chi_{u}(T)
\end{equation}
Here, $\chi_{_{0}}$ is the constant term, which originates from Van-Vleck
paramagnetism and core diamagnetism, $C^{SQUID}$ is the Curie constant. As
discussed in the previous sections, the determination of the superexchange
coupling $J=260\pm10$ K allows us to determine independently $\chi_{u}(T)$ by
using the QMC simulation results, which fit perfectly the mainline NMR
data.\emph{\ }In order to estimate the Curie constant $C$ and $\theta$
reliably, we therefore fitted $\chi_{SQUID}(T)$ for the substituted as well as
the pure samples to Eq. 5 in the temperature range $2$ K $\leq$ $T$ $\leq$
$25$ K. We then fit our susceptibility data ($T<150$ K) for all the samples to
extract the values of $\chi_{_{0}}$ keeping the value of $C$ and $\theta$
obtained from the low-temperature fit. From the fit (shown by solid line in
Fig. 2 (a)), we found that $\chi_{_{0}}$ remains almost unchanged for all the
samples ($\chi_{_{0}}\approx$ $0.3\times10^{-3}$ cm$^{3}$ K/mole). A small
$\theta$ $\approx$\ $1$ K fits the data reliably for all the samples.

\subsection{Comparison of NMR and macroscopic susceptibility in pure YBNO:
nature of intrinsic defects}

Comparison of NMR and SQUID\ measurements in pure non-annealed and annealed
sample allows us to get a better understanding of the various types of defects
present in the \textquotedblleft pure\textquotedblright\ YBNO.\ The annealing
procedure leads to a reduction of the Curie-term $C^{SQUID}$ from
$25.4\times10^{-3}$ cm$^{3}$ K/mole to $13.7\times10^{-3}$ cm$^{3}$
K/mole.\ Such reduction shows that annealing cures part of the defects in the
compound.\ The corresponding\ $^{89}$Y NMR spectra of the non-annealed and
annealed compounds are presented in Fig. 3. In the non-annealed sample,
satellite peaks observed at frequencies smaller than the central peak, are of
higher intensity than those noticed on the higher frequency side. After
annealing, the intensity of the satellites reduced on both sides of the
central peak and the intensity of the left side satellites almost equals to
those in the right side of the mainline. Hence, there are probably various
kinds of defects: in-chain defects which give rise to satellites exactly like
Zn would do, and other defects which lead to the broad low-frequency
feature.\ There must be also magnetic entities at other non-chain sites, which
do not affect the $^{89}$Y NMR signal, since SQUID\ data show a large residual
Curie term in the annealed compound. The in-chain defects could be chain
breaks, due to the polycrystalline nature of the sample, vacancies at Ni-site
and presence of Ni$^{3+}$\ in the chain.\ The other defects could be Ni$^{3+}%
$\ in the interstitial places. Presumably, annealing in vacuum reduces the
quantity of chain breaks (some Ni-ion\ might migrate to the vacant sites on
annealing at $400%
{{}^\circ}%
$C) and Ni$^{3+}$\ in the sample, which is responsible for the reduction of
satellite intensity. Annealing results in a reduction from $0.8$\% to
$0.2$\%/Ni of in-chain defects. If one assumes that each of these in-chain
defects is giving rise, as in the Zn case, to two $S=1/2$\ spins at low-$T$,
the reduction of $0.6$\% percent of defects in the chain by annealing should
lead to a decrease in the Curie constant of $4.5\times10^{-3}$cm$^{3}$K/mole
($0.6\times2\times3.75\times10^{-3}$cm$^{3}$K/mole). Our observed reduction in
Curie constant was higher than this by $7.2\times10^{-3}$cm$^{3}$K/mole. So
the annealing not only reduces the chain defects but might also reduce the
number of Ni$^{3+}$\ (present in YBNO) or remove some of the excess oxygen
present in the other extrinsic paramagnetic phases (say, BaNiO$_{2+x}$). Some
of these defects are however, still present in smaller concentration, even
after annealing.

\subsection{Comparison of NMR and macroscopic susceptibility in YBNO:Zn and
YBNO:Cu: the low-$T$ regime}

A plot of $1/\chi_{SQUID}$ versus temperature done in the insets of Fig. 2,
demonstrates that low temperature susceptibility is dominated by a Curie term,
which increases in the substituted compounds. In Fig. 18, we report the
corresponding Curie constant $C^{SQUID}$ with increasing impurity content for
Cu and Zn. This proves that the substitution induces a paramagnetic
contribution, as already has been shown by NMR at higher temperatures.\ This
contribution seems to persist at temperatures as low as $T=2$ K.\ However, the
\textquotedblleft pure\textquotedblright\ Curie term appears similar in
magnitude to the Zn or Cu contributions for concentrations of about
$0.5$\%.\ A linear behavior of $C^{SQUID}$ with increasing Zn or Cu
concentration is expected. However, we found that the variation of $C^{SQUID}$
with substituent concentration is non-linear and is steeper for higher
concentrations. This limits any straightforward quantitative interpretation of
the data. This limitation is mainly due to the fact that in substituted
samples, a sizable part of $C^{SQUID}$ originates from defects other than the
substitution itself (i.e. background Curie term as discussed in Sec. IVD). The
precise variation of this background Curie term with the substituent
concentration is unknown. It appears that impurity substitution (in small
amounts) at Ni-site helps to decrease this background Curie-term as
$C^{SQUID}$ for Y$_{2}$BaNi$_{0.995}$Cu$_{0.005}$O$_{5}$ is almost identical
to that of the \textquotedblleft pure\textquotedblright\ whereas NMR showed
that Cu induces a paramagnetic contribution. A rough estimate of the impurity
contribution can still be made using the increase of $C^{SQUID}$ for high
enough dopant content. A linear fit (shown by solid lines in Fig. 18) without
taking into\ account the \textquotedblright pure\textquotedblright%
\ composition leads to $C_{Zn}^{SQUID}=(6.4\pm0.7)\times10^{-3}$cm$^{3}$
K/mole/\% Zn and $C_{Cu}^{SQUID}=(3.3\pm0.6)\times10^{-3}$cm$^{3}$%
\textbf{\ }K/mole/\% Cu.

In the Zn case, the NMR$\;$measurements allowed us to extract $C_{Zn}%
^{NMR}=(5.6\pm0.6)\times10^{-3}$ cm$^{3}$ K/mole/\% Zn.\ There is a good
agreement between macroscopic and local measurements even though they apply in
different temperature ranges ($T\leq30$ K and $80$ K$\leq T\leq350$ K
respectively).\ This value is not that far from the VBS simplified picture
where $C_{VBS}=$\ $7.5\times10^{-3}$cm$^{3}$ K/mole\cite{2} originates from
the two chain-end $S=1/2$ spins released by the introduction of Zn.

In the Cu case, if the doped $S=1/2$ Cu acts as a free spin, its
susceptibility should add to that of the induced staggered magnetization
measured by NMR, leading to $\chi_{Cu}^{SQUID}=\chi_{Cu}+\chi_{Cu}^{NMR}(T)$.
Anyhow, we found $\chi_{Cu}^{SQUID}$ to be smaller than $\chi_{Cu}^{NMR}%
=\frac{(4.7\pm0.5)}{T}\times10^{-3}$ cm$^{3}$/mole/\% Cu. This indicates that
the introduced $S=1/2$\ does not act as a free spin, but magnetizes in the
antiparallel orientation to the induced chain-end moments, at least at
low-$T$.\ This result is indeed expected if the Cu interacts
antiferromagnetically with the n.n. chain-end Ni spins, in which case
$\chi_{Cu}<0$\ is expected for $T<J^{\prime}.$\ Such conclusion confirms then
the comparison done previously between QMC\ simulations and our NMR results,
which also leads to an antiferromagnetic coupling between Ni and Cu, with
$J^{\prime}\thickapprox40$ K.

\section{\textbf{CONCLUSION }}

We have probed impurity induced magnetic effects in the Haldane chain compound
YBNO via spinless (Zn) and $S=1/2$\ (Cu) substitutions at the Ni site.\ The
Haldane gap remains unchanged upon substitution. The intensity analysis of the
NMR satellites proves that the amount of dopant impurity present in the sample
agrees well with the nominally intended stoichiometry.\ $^{89}$Y NMR
measurements show that Zn or Cu induce a staggered magnetization in their
vicinity.\ Its magnitude decays exponentially with a characteristic length
equal to the correlation length $\xi$\ of the infinite chain either for non
magnetic or magnetic impurities.\ The fit of the NMR shift data on a large $T$
range for the pure system with the results of QMC\ computations allows us to
deduce a reliable value $J=250\pm10$ K.\ For the Zn impurity the spatial
dependence and magnitude of the staggered susceptibility observed by NMR
matches perfectly the QMC calculated values with $J=270\pm10$ K.\ For $70$ K
$<T<350$ K,\ the total impurity induced Curie susceptibility in the chain
corresponds to an effective spin $S=0.4\pm0.06$\ on either side of the
impurity for the Zn doped case, which is close to the theoretically predicted
$S=1/2$\ excitation at $T=0.\;$

In this high-$T$ range, the induced staggered magnetization is slightly
smaller for Cu which can be explained if the $S=1/2$ Cu spin is coupled
antiferromagnetically with the near neighbor Ni. A fit with the
QMC\ calculated spin polarization allows us to deduce the magnitude
$J^{\prime}=(0.15\pm0.05)J$\ of this AF coupling. This occurrence of an AF
coupling between the Cu and the Ni is confirmed by the fact that the
macroscopic susceptibility induced by Cu as measured by SQUID\ is even smaller
for Cu impurities than for Zn impurities for $T<30$ K.\ This confirms that in
this $T$-range the Cu magnetization opposes that of the local moments of the
chain-ends, as can be expected for an AF\ $J^{\prime}$.

Further experiments are needed to follow the low-$T$\ behavior of the local
magnetization both for Zn and Cu.\ In the case of Zn this would allow to check
whether a Curie law applies to the lowest temperatures, below $T\thickapprox
\Delta$,$\;$as suggested by macroscopic measurements, a matter still debated
theoretically.\ As for the Cu case this would allow us to better follow the
modifications of local magnetization for $T\thickapprox J^{\prime}\;$as strong
deviations from that observed for spinless impurities are expected.\ In case
of high-$T_{c}$ cuprates (e.g. YBa$_{2}$Cu$_{3}$O$_{7} $), substitution of Cu
by magnetic (Ni) or nonmagnetic (Zn) impurity,\cite{d} generates an induced
polarization, with same temperature dependence of the staggered magnetization
in both the cases, but with weaker effects for Ni than Zn.\ This is
qualitatively similar to the present study on a much simpler system, i.e. in
an $1D$ chain, which confirms the validity of the approach which consists in
using magnetic or non-magnetic substitutions in the $2D$ system to probe the
intrinsic magnetic correlations.

\bigskip

\textbf{ACKNOWLEDGEMENT}

Numerical calculations were done using the SSE application package (with
generalized directed loop techniques\cite{alet3}) of the ALPS
project,\cite{alet6} and performed on the Asgard beowulf cluster at ETH Z\"{u}rich.

We thank Fabio Tedoldi, Alberto Rosso, Silvana Botti, Etienne Janod, and Seiji
Miyashita for very fruitful discussions.\ We also would like to thank Indo
French Centre for Promotion of Advanced Research (IFCPAR) for funding this
project (No. 2208-1). F. A. is supported by the Swiss National Science
Foundation, and E. S. S. by the NSERC of Canada and SHARCNET.

\bigskip\newpage

\bigskip\textbf{Figure captions}

FIG. 1: (a) Representation of the unit cell of Y$_{2}$BaNiO$_{5}$ :\ the
$^{89}$Y nucleus is hyperfine-coupled to the Ni electrons in its two next
nearest neighbor (n.n. n) chains through oxygens while its transferred
hyperfine coupling to the Ni on the nearest neighbor (n.n.) chain is expected
to be small. (b) When an impurity is substituted in one of the two n.n. n
chains, $^{89}$Y nuclei which couple to that chain are labelled according to
their distance to the impurity as indicated in the figure. In Fig. 1 (a), only
some of the oxygen atoms are represented, for clarity.\ In Fig. 1 (b), only
the Ni and Y sites are displayed.\ 

FIG. 2: Magnetic susceptibility versus temperature $T$ for (a) Y$_{2}%
$BaNi$_{1-x}$Zn$_{x}$O$_{5}$ and (b) Y$_{2}$BaNi$_{1-x}$Cu$_{x}$O$_{5}$
$(0\leq x\leq0.02)$. The solid line follows Eq. 5 and the fitting was done
using the procedure discussed in Sec. IVC, for the "pure" YBNO sample for $T<$
$150$ K. In the insets, the representation vs $1/T$ evidences the Curie law
behavior at low temperatures. Comparison between non-annealed and annealed
sample in the inset of (a) shows that annealing leads to a reduction of the
Curie term.\ In contrast, Zn or Cu substitution leads to an increase of the
Curie term.

FIG. 3: Comparison of the $^{89}$Y NMR spectra of the non-annealed and
annealed YBNO at $150$ K (in the inset, the full line and in the main frame, a
vertical zoom). The intensity of the satellite peaks reduce after vacuum
annealing showing that the corresponding defects are eliminated by annealing.

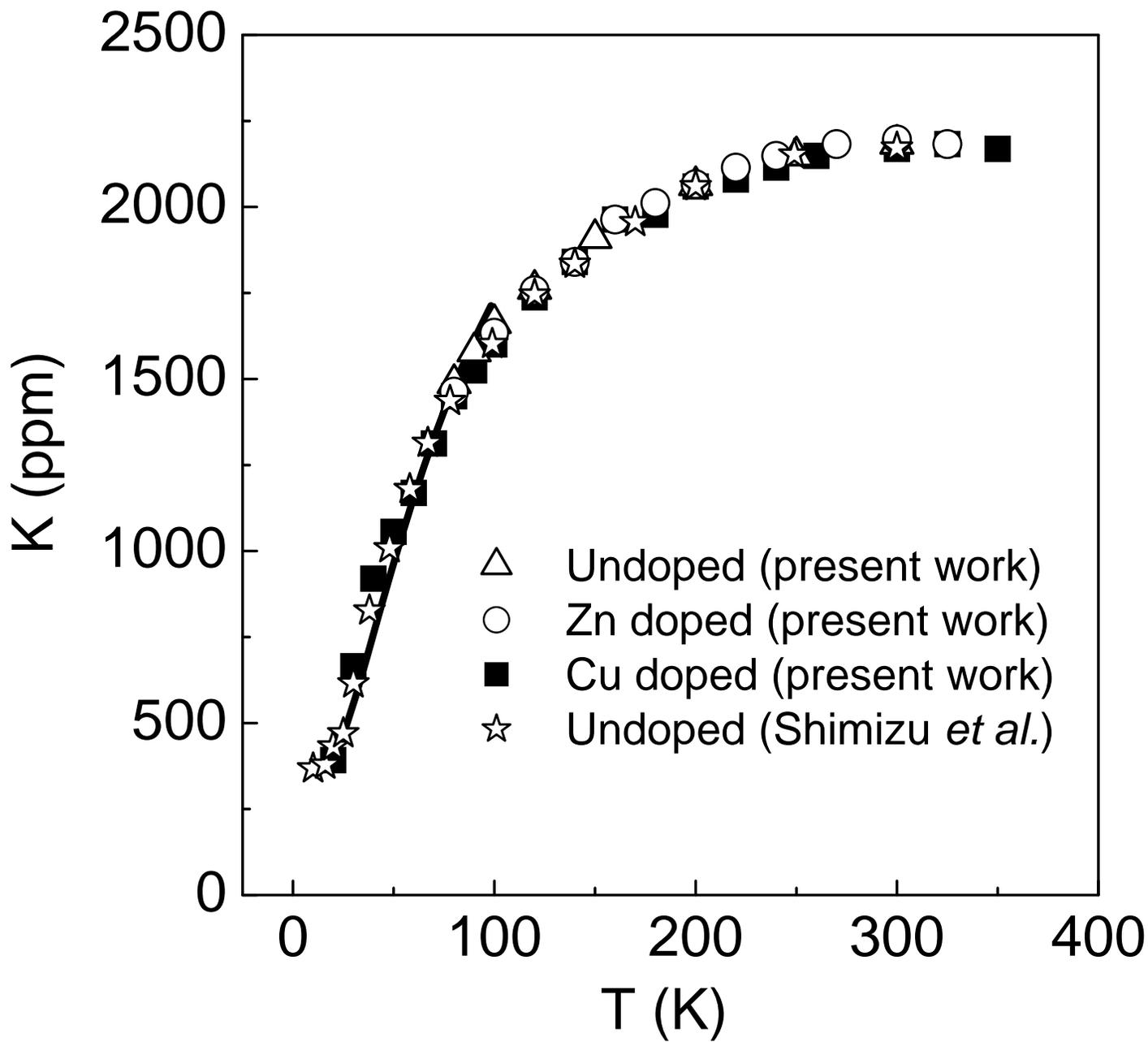
FIG. 4: $^{89}$Y Knight shift of the mainline $^{89}K$ as a function of
temperature $T$, for pure (present results and from Shimizu \textit{et al.}
(Ref. [14])) and substituted samples. $^{89}K(T)$, which measure the uniform
susceptibility, remains unaffected by substitutions. The solid line represents
a fit to Eq. 1 for the Cu doped sample and is discussed in detail in Sec. IVA.

FIG. 5 (a) Expanded view of the satellites of the $^{89}$Y spectra of Y$_{2}%
$BaNi$_{1-x}$Zn$_{x}$O$_{5}$ ($0\leq x\leq0.02$) at $120$ K. Intensity of
these satellites increases with increasing Zn content. The positions of the
satellite peaks are independent of the Zn concentration.(b) Comparison of the
satellites for $0.5$\% and $1$\% Cu substitution at the Ni site at $300$ K.
The satellite intensity increases linearly with Cu content.

FIG. 6: $1^{st}$ and $3^{rd}$ neighbor satellite peaks (labelled '$1$' and
'$3$') for Y$_{2}$BaNi$_{0.98}$Zn$_{0.02}$O$_{5}$ and Y$_{2}$BaNi$_{0.98}%
$Cu$_{0.02}$O$_{5}$\ at $200$ K (solid lines) and fit of satellite $1$ using
an asymmetric Lorenzian lineshape (dotted lines).

FIG. 7: Comparison of $^{89}$Y NMR spectra of Y$_{2}$BaNi$_{0.98}$Zn$_{0.02}%
$O$_{5}$ and Y$_{2}$BaNi$_{0.98}$Cu$_{0.02}$O$_{5}$ measured at $T=200$ K. The
mainline position is the same for both substitutions. The\ satellites for the
Cu substitution are less shifted than those for Zn. Inset: Position of the
satellite peaks on the lower frequency side of the mainline for non-annealed
pure YBNO, Y$_{2}$BaNi$_{0.98}$Zn$_{0.02}$O$_{5}$ and Y$_{2}$BaNi$_{0.98}%
$Cu$_{0.02}$O$_{5}$\ at $200$ K.

FIG. 8: $^{89}$Y NMR spectra of (a) Y$_{2}$BaNi$_{0.98}$Zn$_{0.02}$O$_{5}$ and
(b) Y$_{2}$BaNi$_{0.995}$Cu$_{0.005}$O$_{5}$ at different temperatures.
Satellite peaks are indexed in accordance with their decreasing shift from the
mainline and with the position of the $^{89}$Y\ labelled in Fig. 1 (b). The
peak originating from the Y neighbor to the Zn impurity (as discussed in Sec.
IIIB) is marked by $0$ in (a).

FIG. 9: Temperature dependence of $\delta\nu_{i}$, the frequency shift of the
different satellites with respect to the central line for (a) YBNO:Zn and (b)
YBNO:Cu. The open circles in (a) represent the shifts obtained in Ref. [17]
for YBNO:Mg.\ At lower temperatures, the broadening of the NMR spectra
restricted us from obtaining satellite shifts with high resolution. Indices
represent the $i^{th}$ neighbor to the impurity in accordance with Fig. 1 (b).

FIG. 10: $^{89}$Y\ NMR\ shift of the central line $K$ as a function of
$\chi_{u}$ computed by QMC\ using different $J$ values (with $T$ as an
implicit parameter). The solid line represents the best linear fit and allows
to determine both $J$ and the hyperfine coupling constant $A_{hf}$ (using the
constraint $K(0)=370\pm10$ ppm as detailed in the text).\ 

FIG. 11: Site dependence of the impurity induced magnetic shift $|\delta
\nu_{i}|$ at different temperatures for (a) YBNO:Zn and (b) YBNO:Cu. The
log-scale representation evidences that $|\delta\nu_{i}|$ is exponentially
decaying. The solid lines are fits to the exponential relation represented by
Eq. 2.

FIG. 12:\ Comparison of $\xi_{imp}(T)$\ obtained from our experiments from Eq.
2 and Fig. 11 \ for YBNO:Zn, YBNO:Cu , for YBNO:Mg by Tedoldi \textit{et al.
}(Ref. [17]), and for the infinite chain correlation length $\xi
(T)$\ calculated by QMC\ by Kim et al. (Ref. [33]).

FIG. 13: (a) Comparison at $T=100$ K of the spatial dependence of the
staggered magnetization $\delta<S_{z}^{i}>/<S_{z}>_{u}\;$for YBNO:Zn with the
corresponding QMC\ simulations for various $J$\ values, which shows a good
agreement for $J=270\pm10$ K. (b) Comparison of the $T$\ variation of the
magnetization $<S_{z}^{i}>$ of the $1^{st}$\ and $2^{nd}$\ neighbor ($i=1
$\ and $2$, respectively) with the QMC\ calculations.\ Filled symbols
represent our experimental data while open triangles and circles represent our
QMC results for $J=260$ K. The open stars are the results of classical Monte
Carlo calculations (Ref. [19]).

FIG. 14: Spatial dependence of the experimental ratio $<\delta S_{z}^{i}%
>_{Cu}/<\delta S_{z}^{i}>_{Zn}$of the staggered magnetization near Cu to that
near Zn .\ The data are compared to QMC\ simulations for various values of the
Heisenberg coupling $J^{\prime}$\ between Cu and its n.n. neighbors Ni at
$T=100$ K.

FIG. 15: The impurity induced total susceptibility $\chi_{imp}^{NMR}(T)$
estimated from NMR data are plotted versus $1/T$ for YBNO:Zn$_{1\%}$ (open
triangles) and YBNO:Cu$_{1\%}$ (open squares). The solid lines represents the
Curie-like fits to $\chi_{imp}^{NMR}(T)$. The solid symbols represent the
values obtained at $100$ K from the corresponding QMC\ simulations.

FIG.16: Macroscopic susceptibility of pure YBNO and its decomposition into a
sum of three terms : the QMC-computed (Haldane gap related) decrease of the
uniform susceptibility for $J=260$ K, a Curie term extracted from a fit to the
data for $T<30$ K , and the residual part which cannot be accounted for by the
two preceding forms.

FIG. 17 : The residual term deduced from the macroscopic susceptibility
analysis detailed in the text and in Fig. 16\ are compared here for various samples.

FIG. 18 : Variation of the Curie constant $C^{SQUID}$ extracted from the
low-$T$ susceptibility measurements versus impurity concentration for YBNO:Zn
and YBNO:Cu. The linear fits (solid lines) were done without taking into
account the Curie-term of the \textquotedblleft pure\textquotedblright sample,
as the use of substituent might reduce the concentration of native defects in
the latter.


\bigskip

\begin{thebibliography}{99}                                                                                               %


\bibitem {a}M. Takigawa, N. Motoyama, H. Eisaki, and S. Uchida, Phys. Rev. B
\textbf{55,} 14129 (1997).

\bibitem {13}J. F. DiTusa, S-W, Cheong, J. -H, Park, G. Aeppli, C. Broholm,
and C. T. Chen, Phys. Rev. Lett. \textbf{73,} 1857 (1994).

\bibitem {c}N. Fujiwara, H. Yasuoka, Y. Fujishiro, M. Azuma, and M. Takano,
Phys. Rev. Lett. \textbf{80,} 604 (1998).

\bibitem {d}A. V. Mahajan, H. Alloul, G. Collin, and J. F. Marucco, Phys. Rev.
Lett. \textbf{72,} 3100 (1994); J. Bobroff, H. Alloul, Y. Yoshinari, A. Keren,
P. Mendels, N. Blanchard, G. Collin, and J. F. Marucco, Phys. Rev. Lett.
\textbf{79,} 2117 (1997); J. Bobroff, W. A. MacFarlane, H. Alloul, P. Mendels,
N. Blanchard, G. Collin, and J. F. Marucco, Phys. Rev. Lett. \textbf{83,} 4381
(1999). S. Ouazi, J. Bobroff, H. Alloul, and W. A. MacFarlane, Cond-mat, 0307728.

\bibitem {1}F. D. M. Haldane, Phys. Lett. \textbf{93A}, 464 (1983); Phys. Rev.
Lett. \textbf{50,} 1153 (1983).

\bibitem {2}I. Affleck, T. Kennedy, E. H. Lieb and H. Tasaki, Phys. Rev. Lett.
\textbf{59,} 799 (1987).

\bibitem {3}T. Kennedy, J. Phys. Condens. Matter \textbf{2,} 5737 (1990).

\bibitem {4}D. J. Buttrey, J. D. Sullivan, and A. L. Reingold, J. Solid State
Chem. \textbf{88,} 291 (1990).

\bibitem {5}J. Darriet and L. P. Regnault, Solid State Comm. \textbf{86,} 409 (1993).\ 

\bibitem {b}B. Batlogg, S- W. Cheong, and L. W. Rupp Jr, Physica B
\textbf{194-196}, 173 (1994).

\bibitem {7}T. Yokoo, Y. Sakaguchi, K. Kakurai, and J. Akimitsu, J. Phys. Soc.
Jpn. \textbf{64,} 3651 (1995).

\bibitem {8}Y. Sakaguchi, K. Kakurai, T. Yokoo, and J. Akimitsu, J. Phys. Soc.
Jpn. \textbf{65,} 3025 (1996).

\bibitem {9}G. Xu, J. F. DiTusa, T. Ito, K. Oka, H. Takagi, C. Broholm, and G.
Aeppli, Phys. Rev. B \textbf{54,} R6827 (1996).

\bibitem {10}T. Shimizu, D. E. MacLaughlin, P. C. Hammel, J. D. Thompson, and
S-W. Cheong, Phys. Rev. B \textbf{52,} R9835 (1995).

\bibitem {20}A. P. Ramirez, S- W. Cheong, and M. L. Kaplan, Phys. Rev. Lett.
\textbf{72,} 3108 (1994).

\bibitem {22}C. Payen, E. Janod, K. Schoumacker, C. D. Batista, K. Hallberg,
and A. A. Aligia, Phys. Rev. B \textbf{62,} 2998 (2000).

\bibitem {24}F. Tedoldi, R. Santachiara, and M. Horvati\'{c}, Phys. Rev. Lett.
\textbf{83,} 412 (1999).

\bibitem {23}C. D. Batista, K. Hallberg, and A. A. Aligia, Phys. Rev. B
\textbf{58,} 9248 (1998).

\bibitem {25}S. Botti, A. Rosso, R. Santachiara, and F. Tedoldi, Phys. Rev. B
\textbf{63,} 012409 (2001).

\bibitem {26}S. Miyashita and S. Yamamoto, Phys. Rev. B \textbf{48,} 913 (1993).

\bibitem {29}F. Alet and E. S. S\o rensen, Phys. Rev. B \textbf{62,} 14116 (2000).

\bibitem {27}E. S. S\o rensen and I. Affleck, Phys. Rev. B \textbf{51,} 16115 (1995).

\bibitem {28}E. Polizzi, F. Mila, and E. S. S\o rensen, Phys. Rev. B
\textbf{58,} 2407 (1998).

\bibitem {30}R. D. Shannon and C. T. Prewitt, Acta Cryst. \textbf{B 25,} 925 (1969).

\bibitem {alet1}A.W. Sandvik and J. Kurkij\"{a}rvi, Phys. Rev. B \textbf{43},
5950 (1991).

\bibitem {alet2}A. W. Sandvik, J. Phys. A \textbf{25}, 3667 (1992).

\bibitem {alet3}F. Alet, S. Wessel and M. Troyer, cond-mat/0308495.

\bibitem {alet4}O. F. Sylju\"{a}sen and A. W. Sandvik, Phys. Rev. E
\textbf{66}, 046701 (2002).

\bibitem {alet5}O. F. Sylju\"{a}sen, Phys. Rev. E \textbf{67}, 046701 (2003).

\bibitem {12}F. Tedoldi, A. Rigamonti, C. Brugna, M. Corti, A. Lascialfari, D.
Capsoni, and V. Massarotti, J. App. Phys. \textbf{83,} 6605 (1998).

\bibitem {AffleckA}I.\ Affleck, Phys.\ Rev.\ B \textbf{41}, 6697 (1990); see
also private communication by I.\ Affleck reported in M. Takigawa, T. Asano,
Y. Ajiro, and M. Mekata, Phys. Rev. B \textbf{52}, R13087 (1995).

\bibitem {31}T. Ami, M. K. Crawford, R. L. Harlow, Z. R. Wang, D. C. Johnston,
Q. Huang, and R. W. Erwin, Phys. Rev. B \textbf{51,} 5994 (1995).

\bibitem {33}Y. J. Kim, M. Greven, U. J. Wiese, and R. J. Birgeneau, Eur.
Phys. J. B \textbf{4,} 291 (1998).

\bibitem {37}P. Roos and S. Miyashita, Phys. Rev. B \textbf{59,} 13782 (1999).

\bibitem {38}T. Tonegawa and M. Karubaki, J. Phys. Soc. Jpn. \textbf{64,} 3956 (1995).

\bibitem {alet6}See http://alps.comp-phys.org.
\end{thebibliography}
\end{document}